\documentclass[preprint,12pt]{elsarticle}

\usepackage{amssymb}
\usepackage{subfigure}
\usepackage{balance}
\usepackage{graphics}
\usepackage{url}
\usepackage{enumitem}
\usepackage{colortbl}
\usepackage{tabularx}
\usepackage{graphicx}
\usepackage{pseudocode}
\usepackage[figuresright]{rotating}
\usepackage{changepage}
\usepackage{gensymb}
\usepackage{microtype,amsmath}
\usepackage{hhline}
\usepackage{datetime}
\usepackage{stfloats}
\usepackage{multirow}
\usepackage{booktabs}
\usepackage{longtable}
\usepackage{soul}
\usepackage{color, colortbl}
\usepackage{setspace}
\usepackage{pdflscape}

\newcommand{\revision}[1]{\textcolor{red}{#1}}

\journal{}

\begin{document}

\begin{frontmatter}



\title{Cognitive Performance Measurements and the Impact of Sleep Quality Using Wearable and Mobile Sensors}


\author[inst1,*]{Aku Visuri}
\author[inst2]{Heli Koskimäki}
\author[inst3]{Niels van Berkel}
\author[inst1]{Andy Alorwu}
\author[inst1]{Ella Peltonen}
\author[inst4]{Saeed Abdullah}
\author[inst1]{Simo Hosio}
\affiliation[inst1]{organization={Center for Ubiquitous Computing, University of Oulu},
            addressline={Pentti Kaiteran katu 1}, 
            city={Oulu},
            postcode={90570}, 
            state={Oulu},
            country={Finland}}

\affiliation[inst2]{organization={Oura Research, Oura Health Ltd.},
            addressline={Elektroniikkatie 10}, 
            city={Oulu},
            postcode={90590}, 
            state={Oulu},
            country={Finland}}
\affiliation[inst3]{organization={Human-Centred Computing group, Aalborg University},
            addressline={Selma Lagerløfs Vej 300}, 
            city={Aalborg},
            postcode={9220}, 
            state={Aalborg},
            country={Denmark}}
            
\affiliation[inst4]{organization={Wellbeing \& Health Innovation lab, Penn State},
            addressline={E397 Westgate Building}, 
            city={University Park},
            postcode={16802}, 
            state={PA},
            country={United States}}

\affiliation[*]{Corresponding Author}{}

\begin{abstract}
Human cognitive performance is an underlying factor in most of our daily lives, and numerous factors influence cognitive performance. In this work, we investigate how changes in sleep quality influence cognitive performance, measured from a dataset collected during a 2-month field study. We collected cognitive performance data (alertness) with the Psychomotor Vigilance Task (PVT), mobile keyboard typing metrics from participants' smartphones, and sleep quality metrics through a wearable sleep tracking ring. Our findings highlight that specific sleep metrics like night-time heart rate, sleep latency, sleep timing, sleep restfulness, and overall sleep quantity significantly influence cognitive performance. To strengthen the current research on cognitive measurements, we introduce smartphone typing metrics as a proxy or a complementary method for continuous passive measurement of cognitive performance. Together, our findings contribute to ubiquitous computing via a longitudinal case study with a novel wearable device, the resulting findings on the association between sleep and cognitive function, and the introduction of smartphone keyboard typing as a proxy of cognitive function. 
\end{abstract}


%


\end{frontmatter}


\section{Introduction}
Our cognitive performance varies over time and is influenced by many factors in our daily life. In the context of computer systems, continuously assessing changes in cognitive performance can lead to a better understanding of user needs and more personalised systems. In this work, we first establish how our cognitive performance changes and is influenced by a range of factors, in our case, how it is influenced by sleep quality. Sleep is one of the fundamental driving forces in our daily lives, daily performance, and overall well-being. Sleep affects various aspects of our overall life quality \cite{bonnet1985effect}, behaviour, and physiology \cite{bonnet1989effect}. Modern sleep tracking devices have significantly evolved in the last few years, offering new accurate tools for long-term and convenient sleep monitoring \cite{de2019sleep, de2018validation}. With these devices, researchers have access to more fine-grained sleep metrics, e.g., sleep stages, heart rate, and restfulness, which can help uncover previously unknown relationships between human behaviour performance and quality of sleep. In this paper, we focus on the impact of sleep on cognitive performance, which is imperative in task and work effectiveness \cite{kleitman1923studies,miyata2013poor,scullin2015sleep,alhola2007sleep,pilcher1997sleep}.

The works of, e.g., Abdelrahman et al. \cite{abdelrahman2017cognitive} and Abdullah et al. \cite{abdullah2016cognitive} have taken steps in recent years for creating unobtrusive and continuous methods of tracking alertness, which is an indicator of cognitive load and performance. These tools and methods could prove helpful in the future for creating cognition-aware software. This paper contributes to this area by demonstrating how passive sensing of smartphone typing can be utilised to evaluate changes in alertness. We explore how smartphone typing metrics can indicate changes in cognitive performance. Given the integration of smartphones into our day-to-day lives, these metrics can lead to an unobtrusive, continuous, and in-situ assessment of cognitive performance. Cognition-aware systems have gained attention in the ubiquitous research community, enabling systems to adjust their behaviour (adjust interfaces, prevent disruptions, trigger interventions, etc.) according to the user's cognitive state \cite{mathur2017towards}. 

This paper uses detailed modern sleep-tracking devices with smartphone sensing and a cognitive performance measurement toolkit. We validate the used approach by measuring the previously known effect of sleep quality on alertness and exploring new ways to capture such aspects of cognitive performance unobtrusively. We present the results of a 2-month study designed to capture i) sleep quality through a wearable ring sensor, ii) smartphone typing metrics through the participant's smartphone, and iii) the participant's alertness levels through a mobile application that measures alertness with the Psychomotor Vigilance Task (PVT) test. The PVT test is one of the gold standards for measuring alertness, but it suffers from a lack of continuous tracking and the requirement of active participation. To look for alternative methods for passively sensing alertness, we look into typing metrics, similar to Althoff et al. \cite{alhola2007sleep}. Our results in Chapter \ref{results3} showcase our findings on how the interaction between cognitive performance, sleep quality, and smartphone typing data can be transitioned from one to another; sleep quality influences both alertness and typing metrics similarly, and changes in alertness (e.g., worse reaction times) correlate with typing metrics (e.g., slower typing pace). The contributions of this paper are as follows:
\begin{itemize}
    \item We present a longitudinal study of two months, where ubiquitous sensing technologies were used to collect sleep and typing data for the full study duration and reaction test data for two weeks. 
    \item We show the impact of individual sleep metrics on cognitive performance, highlighting that various factors must be considered when evaluating the effect of sleep and that modern digital sleep tracking devices can offer previously unrevealed insights about the effect of sleep. 
    \item Using the passively sensed datasets from our study, we investigate and showcase how smartphone keyboard typing metrics could function as a novel proxy for existing cognitive performance measurement techniques or as a complementary method to both active and passive cognitive performance measurement devices.
\end{itemize}

\section{Related Work}
We present existing literature on cognitive performance measurements, sleep-tracking technologies, and how smartphone usage characteristics have been used to measure human behaviour. First, we briefly touch on how cognitive performance fluctuates and is measured, then the topic of sleep research and what type of influence sleep is shown to have on cognitive performance. Lastly, we provide an overview of what types of insights can be uncovered from analysing one's typing habits and actions. 

\subsection{Measuring Cognitive Performance}
Natural biochemical processes influence both physical and cognitive performance. Our body goes through diurnal variations throughout the night and day, influencing our physical and cognitive performance. Kleitman \cite{kleitman1923studies} was one of the initial finders of these factors. Borbély and Alexander \cite{borbely1982two} attribute these variations to two underlying processes: 1) the homeostatic process and 2) circadian variations. The homeostatic sleep pressure builds up as we spend time awake and typically causes our performance to decline over time.

Meanwhile, the circadian process manifests in a sinusoidal pattern, which determines when we typically experience highs and lows during the day in both sleep drive and performance. Due to the nature and shape of the two processes, we typically peak in performance in the afternoon \cite{van2000circadian}. Both processes influence our perception, memory capabilities, decision-making, and overall cognitive performance \cite{o2013impact,schmidt2007time,thomas2000neural}. When the processes cause our performance to decline, we are at higher risk of making mistakes \cite{blatter2007circadian,hanecke1998accident}, fatigue-induced accidents \cite{dinges1995overview} and generally suffer from lowered alertness. Additionally, poor sleep has been shown to impair cognitive performance \cite{miyata2013poor}, while good sleep quality promotes better cognitive function and protects against age-related cognitive declines \cite{scullin2015sleep}.

Systems that analyse these rhythms and performance are referred to as cognition-aware systems \cite{mathur2017towards}. Different cognitive states can be used as one factor of the user's context to appropriately schedule work tasks throughout the day or inform us about our current potential. Cognition-aware systems can adjust work tasks, timings, and execution times or even adjust interfaces in their complexity according to the user's current capabilities \cite{bulling2014cognition,dingler2016cognition}. The users' alertness levels could be measured for these systems to function properly and more effectively. Traditionally used methods (e.g., core temperature or cortisol and melatonin level measurements) are typically quite invasive, as are lengthy observation periods in sleep labs \cite{hofstra2008assess}. Less intrusive methods consist of, e.g., alertness self-assessments like the Karolinska Sleepiness Scale (KSS), through which users rate their state of drowsiness \cite{aakerstedt1990subjective}. An alternative option is the Stanford Sleepiness Scale (SSS) \cite{hoddes1973quantification}. Self-reports typically suffer from reliability issues due to the certain degree of subjectivity of the self-reflection and introspection required \cite{fairclough2009fundamentals,van2000circadian}.

Better alternatives consist of objective measures, such as alertness assessment tasks, which typically measure reaction times or accuracy. Methods for continuous alertness assessment methods have been explored by e.g., Tag et al. \cite{tag2019continuous}, with the use of EOG glasses and with the use of mobile games by Dingler et al. \cite{dingler2020extracting}. Abdullah \cite{abdullah2015towards} coined the term circadian computing, which aims to detect and provide in-situ interventions related to our daily biological rhythms. The Psychomotor Vigilance Task (PVT) is commonly used to assess reaction times to randomly appearing visual stimuli \cite{dinges1985microcomputer}. The PVT test is easy to deploy on users' mobile phones \cite{kay2013pvt} or integrate into more extensive cognitive assessment toolkits \cite{dingler2017building}. These methods have been used to assess the alertness levels of study subjects, for example, to study the association between smartphone usage behaviour and corresponding reaction times throughout the day \cite{abdullah2016cognitive}. Other cognitive assessment tasks consist of the Go-NoGo (GNG) \cite{dingler2017building} and the Multiple Object Tracking (MOT) \cite{pylyshyn1988tracking}. GNG includes an assessment of decision-making abilities, and MOT measures the ability to focus on multiple stimuli in a split-attention scenario. 

Dingler et al. \cite{dingler2017building} validated these three standardised tasks and their ability to detect changes in homeostasis and circadian rhythms. They also published their cognitive toolkit as an open-source library. Thus, we have opted to use their library in our study setting. We opt to only use the PVT test according to Dingler et al. \cite{dingler2017building} as it is the only one of the three that simultaneously tracks homeostatic pressure and circadian rhythm changes. 

\subsection{Analysing Sleep Quantity and Quality}

Sleep is typically measured by quantity ("how much sleep you got?") and quality ("how restful your sleep was?"). Traditional sleep research is separated into two main approaches:  physiological measurements and self-report methods. The PSG is an objective measurement done primarily in clinical settings which classifies sleep into 30-second epochs of different sleep stages (phases) ~\cite{bloch1997polysomnography,chesson1997indications,buysse2014sleep} based on AASM (American Association of Sleep Medicine) standards. The stages used are wake, slow-wave sleep (SWS, also known as N3), N2, N1 and REM sleep. Other metrics include sleep latency, wake after sleep onset (WASO) and amount of time in sleep, i.e. sleep efficiency. PSG is typically used to diagnose sleep disorders like sleep apnea, insomnia, restless legs syndrome, and parasomnias. PSG is an objective measurement done primarily in clinical settings.

Longer observational studies that intend to reveal population-wide sleep results often rely on self-reported metrics through sleep questionnaires and diaries \cite{ibanez2018survey}. Although subjective measures are less accurate than objective measures, objective measures like PSG tend to lack the usability or comfort required for long-term tracking \cite{ibanez2018survey}.

To complement the traditional sleep research and to bring sleep information to everyday users, there is currently rapid growth in the development of sleep-tracking applications and devices. Wearable devices augmented with PPG-based blood flow measurement, heart rate and respiratory rate estimation classify sleep stages into four classes (deep, light, REM, wake), where deep sleep term is used for slow-wave sleep, and light sleep contains both N1 and N2 sleep stages. Although these approaches are validated to correlate with the PSG sleep stages ~\cite{de2019sleep,de2018validation}, some clinicians have remained sceptical of the accuracy and use of sleep information offered by consumer devices for clinical use~\cite{de2016boom,shelgikar2016sleep}. Kuosmanen et al. analysed commercial wearable devices and highlighted significant differences between devices in tracking sleep stages \cite{kuosmanencomparing}.

In summary, sleep has been shown to affect both the physiological and mental aspects of human life. Modern technological solutions offer new ways to accurately and continuously track sleep during our everyday lives. Typically used sleep tracking methods and their attributes are summarised in Table \ref{tab:sleep_tracking_methods}. We also include examples of smartphone applications like Sleep as Android\footnote{\textit{Sleep as Android}, https://play.google.com/store/apps/details?id=com.urbandroid.sleep}, which leverage user input or smartphone sensors to approximate sleep times as one method, albeit these are not commonly used in research contexts. 
 
\begin{table}[t]
\tiny
\begin{tabular}{lcc|ccc|c}
\multicolumn{1}{c}{}   & & &  & \textit{Tracked Variables} & &  \\
\multicolumn{1}{l}{\textbf{Tracking method}} & 
  \begin{tabular}[c]{@{}l@{}}\textbf{Data} \\ \textbf{collection} \end{tabular} &
  \textbf{Accuracy} &
  \begin{tabular}[c]{@{}l@{}}\textbf{Sleep} \\ \textbf{phases} \end{tabular} &
 \begin{tabular}[c]{@{}l@{}}\textbf{Sleep} \\ \textbf{duration} \end{tabular} &
  \textbf{Other} \textbf{metrics} &
  \begin{tabular}[c]{@{}l@{}}\textbf{Enables} \\ \textbf{continuous} \\ \textbf{tracking}\end{tabular} \\ \hline
Polysomnography        & Objective  & High       & Yes               & Yes           & Sleep interruptions & No  \\
Survey or self-report  & Subjective & Low-Med & No                & Yes (approx.) & -                   & Yes \\
Smartphone application & Objective  & Low        & No                & Yes (approx.) & -                   & Yes \\
Wearable tracker &
  Objective &
  Med-High &
  Yes &
  Yes &
  \begin{tabular}[c]{@{}l@{}}Heart rate, temperature,\\ sleep interruptions\end{tabular} &
  Yes
\end{tabular}
\caption{Typically measured sleep tracking methods and their attributes}
\label{tab:sleep_tracking_methods}
\end{table}

\subsection{Typing Analytics}
Smartphone use is frequently used in Human-Computer Interaction studies to better understand human behaviour through interactions with the smartphone \cite{van2016systematic}, responses to interruptions \cite{visuri2017predicting}, biological processes \cite{murnane2015social,murnane2016mobile}, mental states like stress \cite{bauer2012can,vizer2009automated}, and depression \cite{opoku2019towards}. Large-scale smartphone interaction data have been used to gain insights into human behaviour in the areas of mood rhythms \cite{golder2011diurnal}, diet \cite{west2013cookies}, conversation strategies \cite{althoff2016large}, social networks and mobile games encouraging health behaviours \cite{althoff2017online,althoff2016influence,shameli2017gamification}, and health and disease-related search behaviours \cite{paparrizos2016screening,white2016early}. Keystroke patterns, in particular, have largely been left to be used in user authentication, a research focus that began in the 80s and has frequently been revisited since \cite{bergadano2002user,gaines1980authentication}. The emergence of smartphones started to shift the focus from desktop typing to smartphone typing but mainly remained in the same area of user authentication \cite{kumar2016continuous,buriro2015touchstroke}. Keystroke logging does have other applications, such as identifying writing strategies and understanding cognitive processes through typing patterns \cite{deusing} and when pauses occur \cite{leijten2019analysing}. Typing analytics have also been used to measure stress \cite{vizer2009detecting,saugbacs2020stress} and human emotions \cite{ghosh2017tapsense}. The latest works expand on analysing how predictive text entry affects human emotion \cite{ghosh2019does}.

Typing analytics as a performance metric has only been briefly studied as an analysis method for cognitive performance or the impact of sleep on performance. Althoff et al. \cite{althoff2017harnessing} presented large-scale results on how desktop search-engine typing is influenced by circadian rhythms and sleep quality. Our works introduce a similar approach but leverage smartphone typing - a ubiquitous typing activity that requires a different set of nimbleness and occurs in more varying situational contexts. Vizer et al. \cite{vizer2009detecting} proposed an approach to detect cognitive and physical stress from typing metrics, and Mastoras et al. \cite{mastoras2019touchscreen} used touchscreen typing as a means to detect depressive tendencies.

Previous literature has shown how sleep impacts our physiological and cognitive performance and how these can be measured in laboratory settings and daily life. Behavioural metrics collected through smartphone sensing have also been shown to be associated with behavioural tendencies, biological processes, and mental states in human beings. Next, we will describe our study design and how we collect information about our study subjects' sleep, cognitive performance, and smartphone usage behaviour.

\section{Study Design}

\subsection{Apparatus and Data Collection}
We designed a 2-month-long experiment. We used a wearable ring sensor manufactured by Oura\footnote{https://www.ouraring.com} \cite{koskimaki2018we}, which includes a variety of sensors, including infrared photoplethysmography (PPG) for heart rate, negative temperature coefficient (NTC) for body temperature, and a 3D accelerometer. It measures daily activity and performance as well as detailed sleep-related metrics. We also collected smartphone usage behaviour from the participants' devices using a bespoke software application during the study period. Finally, we collected participants' alertness levels multiple times per day using the PVT test during the final two weeks of the study period. The two-month study period was selected as an appropriate length to apply relatively long-term tracking that would reduce the impact of the novelty effect \cite{shin2019beyond} but would not be considered too burdensome by the study participants. Participants responded to a preliminary questionnaire about their demographic information and the standardised MEQ (morningness-eveningness) questionnaire about their circadian cycle. The MEQ results did not yield any significant results and were not included in the analysis of this work.

\begin{table}
\centering
\small
\begin{tabular}{ll}
\textbf{Metric}                & \textbf{Description}                                                                                            \\ \hline
Total time in bed     & Total time (in minutes) in bed, from going to bed to getting up\\
Total time of sleep   & Total duration (in minutes) of sleep                                                                   \\
Sleep latency         & Time (in minutes) it took to fall asleep on the onset of bedtime                                       \\
Sleep efficiency      & Time (in percentages) spent asleep during bedtime                                                      \\
Sleep phases          & Time (in minutes) spent in different sleep phases (light, deep, REM, awake)                            \\
Night-time HR \& HRV & Resting night-time heart rate, heart rate variability      \\
Restfulness           & Percentage of restless events during the night (lower is better)           \\
Sleep score           & Aggregated sum of sleep rates variables to a score (1-100)
\end{tabular}
\caption{Sleep metrics provided by the wearable ring through the vendor's cloud service.}
\label{tab:oura}
\end{table}

\begin{figure}[t]
    \centering
    \includegraphics[width=0.15\columnwidth]{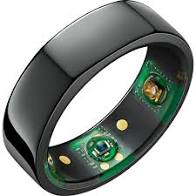}
    \caption{The Oura sleep tracking ring. The sensors - infrared photoplethysmography (PPG) for heart rate, negative temperature coefficient (NTC) for temperature, and 3D accelerometer for movement - are placed on the inner surface of the ring.}
    \label{fig:ouraring}
\end{figure}

\subsubsection{Sleep Data}
Sleep data for our study was provided by the Oura Ring. The ring's form factor is inconspicuous or even fashionable (according to user anecdotes), and with an average battery life of 3-5 days, it enables continuous tracking. 
The accuracy of the tracking of sleep phases is closely related to analysis from polysomnography devices \cite{de2019sleep}, and the ring can accurately measure resting heart rate and heart rate variability \cite{kinnunen2020feasible,stone2021assessing}. According to comparison to a PSG device\cite{de2019sleep}, the Oura ring is accurate at measuring sleep onset, total sleep time, and wake time. Sensitivity to measure different sleep phases was between 51\% (deep sleep) to 65\% (light sleep). This inaccuracy should be considered as we present the analysis regarding different sleep phases. Oura can underestimate the duration of deep sleep (up to 20 minutes) and overestimate REM sleep (up to 17 minutes).
Table \ref{tab:oura} shows a summary of the sleep-related metrics collected through the wearable ring, and Figure \ref{fig:ouraring} shows the form factor and placement of sensors.

\subsubsection{Cognitive Performance}

Within the two-month study period for each participant, we devoted the study's final two weeks (14 days) to collecting reaction test data through a mobile PVT tool called Circog. Circog was created for research purposes by Dingler et al. \cite{dingler2017building}. We modified the existing open source library, which is available in GitHub\footnote{https://github.com/Til-D/circog} and altered the application to suit our needs; we use an experience sampling method (ESM) to deliver notifications to the participants through the application at random 1-3 hour intervals between 9 am and 10 pm. Additionally, a reminder notification is sent after the participant has actively used his or her smartphone for longer than 15 minutes, as provided as an efficient guideline in \cite{visuri2017challenges}. 

\begin{figure}[h]
    \centering
    \includegraphics[width=0.8\columnwidth]{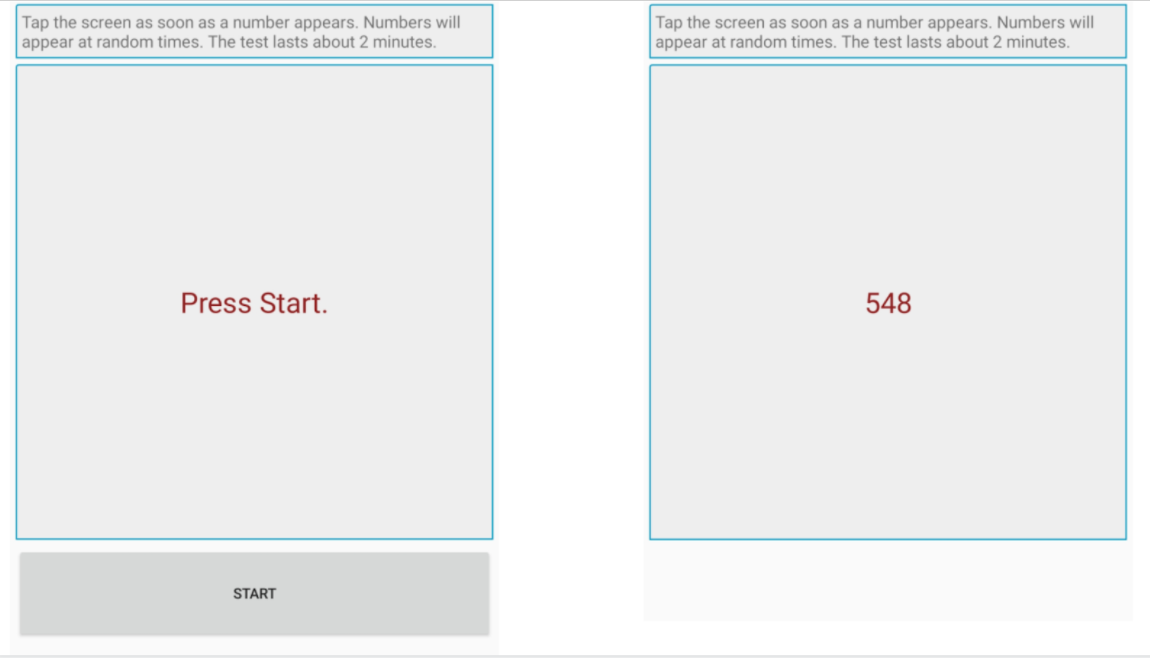}
    \caption{The reaction test application ("Circog") used in our study. Once the user clicks start (left figure), at random intervals the counter (right figure) starts running after 3-10s. Clicking anywhere on the surrounding box stops the counter, stores the reaction time result and starts a new counter after a random 1-5s interval. This is repeated 8-12 times. Details of the application can be found in \cite{dingler2017building}.}
    \label{fig:circog}
\end{figure}

Clicking on the notification launches Circog's PVT test task, which takes approximately 1-2 minutes to complete. The test task consists of 8-12 (randomised) reaction tests where a counter starts running at random intervals (3-10 seconds after the prior has finished), and the participant must click anywhere on the screen as soon as possible. Figure \ref{fig:circog} shows a view of the test task. If the participant clicks too soon (before the counter appears), that test is flagged as an error. The application logs all reaction times during a task, the number of errors, and when the test task started. We use Google Firebase Real-time database\footnote{https://firebase.google.com/} to store the test results in the JSON format anonymously. \revision{An example of data is shown in Table \ref{tab:pvt_data}.}

\begin{table}
\centering
\begin{tabular}{l|l}
\textbf{Variable name} & \textbf{Example value}                                \\ 
\hline
startTasksTime         & 1577809895909                                         \\
endTasksTime           & 1577809944872                                         \\
measurements           & {[}312, 339, 324, 297, 310, 337, 281, 317, 282, 295]  \\
numberOfErrors         & 0                                                     \\
taskCompleted          & true                                                  \\
device\_id               & acc8c1a5420aed8b                                      \\
email                  & $<$participant\_anonymised\_identifier$>$              
\end{tabular}
\caption{Example of the data collected from a PVT test task.}
\label{tab:pvt_data}
\end{table}

The standardized PVT is conducted over ten minutes, but we have consciously decided to adapt to a shorter version. ESM-based studies typically lack participant retention, especially if the conducted tasks are too laborious. Recent research findings support increased burden and compromised data quantity and quality with longer questionnaires, but not with increased sampling frequency \cite{eisele2022effects}. A full 10-minute or 5-minute PVT conducted multiple times per day between daily tasks would likely compromise data quantity. Research into shorter-form PVTs has shown that while a shorter 2-minute version is not as sensitive in measuring sleep loss, the shorter version still strongly correlates with the long version (r = .77) \cite{roach2006can}.

Participants can ignore notifications but were requested to attempt to complete four daily test tasks on each of the 14 days. They could also finish test tasks by simply opening the Circog application. To measure participants' cognitive load, we only considered participants who provided enough (at least two) reaction test tasks on at least fourteen days. 

\subsubsection{Smartphone Keyboard Typing Data}

Smartphone usage data is collected through the AWARE framework \cite{ferreira2015aware}, used extensively in similar HCI research. AWARE enables background logging of different datasets from the user's smartphone, which are anonymously stored on a remote server. For the purposes of this article, we collected typing information through the keyboard sensor \footnote{https://awareframework.com/keyboard/}. The keyboard sensor collects typed characters (which are obfuscated to lower or upper capital a-letters to retain the participants' privacy), the timestamp of the typing event in milliseconds, the application package name where the typing event occurred, and what was the typed text before and after each keyboard event (e.g., typing a character). This enables the sensor to recognise backspaces and other corrective behaviours.

\subsection{Participant Recruitment}
Participants were attracted through mailing lists and word of mouth. Participants received no monetary awards or other compensation for participation. Participants received a sleep-tracking ring to be used during the study period and an Android smartphone in cases when they did not possess one or had a model incompatible with the used tracking applications. Each device was returned to the researchers after the study period was concluded. Our initial call attracted 452 potential participants. We selected two batches of 50 participants (totalling up to 100) to be invited to the study, counterbalancing for age and gender.

We sent out invitations for the selected 100 participants to meet with the researchers, consent to the study procedures, install the required logging applications and measure their finger size for the ring. The delivery of the rings lasted approximately two weeks, after which the participants met with the researchers for a second time, received the ring, and resolved any problems with the logging applications. The two batches of participants were managed back to back, with a short period in between to conclude the study for the first group and contact the invited participants for the second group.

Participants were invited to start the study in person, either one-on-one or in a small group of five or fewer (participants' choice) with one of the research team members. In the initial meeting, the participants studied and signed a consent form, were given a participant ID and a pre-generated email address for enabling pseudonymous data collection, and received written and oral instructions on how to get started with the tracking applications. The participants were also encouraged to check their emails periodically, just in case there was a need for any actions to be taken during the study. Each participant's data was remotely monitored to capture any issues in the data collection. Any participant with a data collection issue could then be contacted via email to mitigate any future issues. At the end of the study, the participants returned their borrowed devices and partook in an end-of-study interview, either in person or through an online survey. There was no enforcement regarding the ring use; participants were asked to wear it as it best suited them during their everyday lives. 

\subsection{Ethical Considerations and COVID-19 Restrictions}
The study design and data collection were subject to the IRB process of the host university. As there was no particular intervention in the study, the local board did not require a full ethics permit process. Instead, in the initial meeting, the participants were informed of their consent and signed a consent form sanctioned by the ethical board of the host university. Participation in the study was voluntary, and the users were informed about the data collection and management procedures.

We also note that the study partially overlapped with the COVID-19 pandemic, causing altercations to the original study plan. Particularly due to the social distancing guidelines of our host organisation, the second participant group were not asked to return their devices in person at the end of their designated two-month period. For these participants, the data collection phase continued for as long as the participant opted to continue the use of the wearable. The participants were asked to respond to an end-of-study online survey that covered similar topics to the interview session for the first participant group. For this study, we only use the data collected during the original length two-month study period. 

\begin{figure}[h]
    \centering
    \includegraphics[width=1\columnwidth]{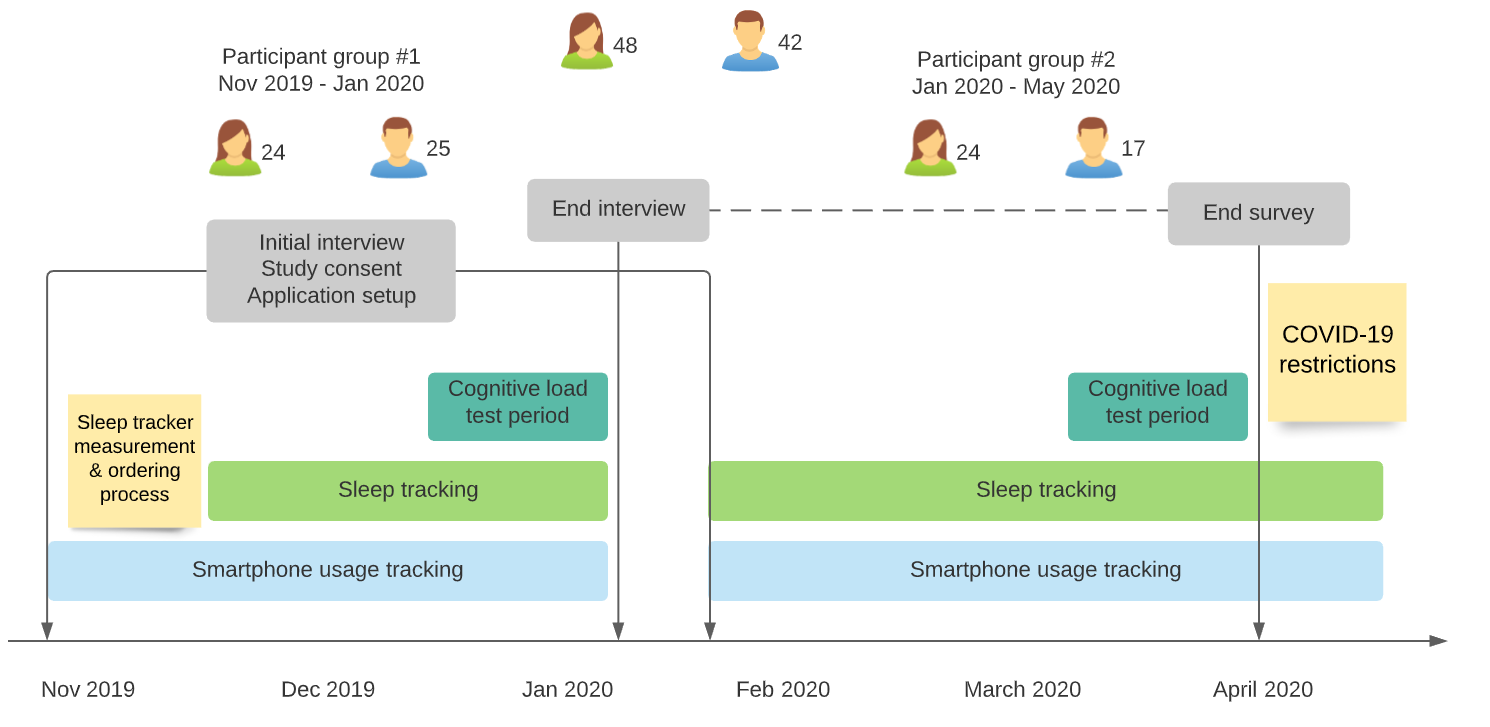}
    \caption{Study periods and participants of both participant groups in the study. For participant group 2 the end interview was replaced with an online survey with same topics being addressed, due to the COVID-19 social distancing restrictions. Group 2 could re-use sleep trackers from group 1, thus for most their sleep tracking period begun right away.}
    \label{fig:study_design}
\end{figure}

\section{Data Analysis}

\subsection{Overview and Data Cleansing}

Initially, we selected 50 participants for each participant group (Group 1 and Group 2), of which 49 joined Group 1 (November 2019) and 41 joined Group 2 (January 2020), totalling 90 participants. Due to researcher time constraints, we did not replace unresponsive or no-show participants as the study schedule was densely packed. Out of the 90, two participants dropped out mid-study; one had a skin condition preventing wearing the wearable, and one did not want to borrow an Android device for the study. In addition, two participants did not opt-in to share their sleep data through the wearable vendor's cloud service. Figure \ref{fig:study_design} illustrates the study periods, phases, and the number of participants. Some of the ring sizes could be reused from Group 1 participants to Group 2, which allowed several Group 2 participants to use the wearable directly after their initial meeting with the researcher(s).

\subsubsection{Sleep Data}
The collected data is of high quality, with only a few gaps in the sleep data. Typically, our sleep dataset's gaps last 1-2 nights. 86 participants provided their sleep data on a total of 3764 unique nights. Of these, 48 participants (53.3\%) recognised as female and 42 (46.6\%) as male. The mean age of participants was 27.0 years (median of 24 years), with a standard deviation of 7.61 years, minimum age of 18 and maximum age of 61 years. All participants were either students or employed at the local university. Participants shared an average of 43.76 nights of sleep data with a standard deviation of 10.5. The largest sample comes from a participant who shared 68 nights of sleep data. The student participants were likelier to have abnormal sleep routines, especially during spring break.

\begin{figure}[t]
    \centering
    \includegraphics[width=0.7\columnwidth]{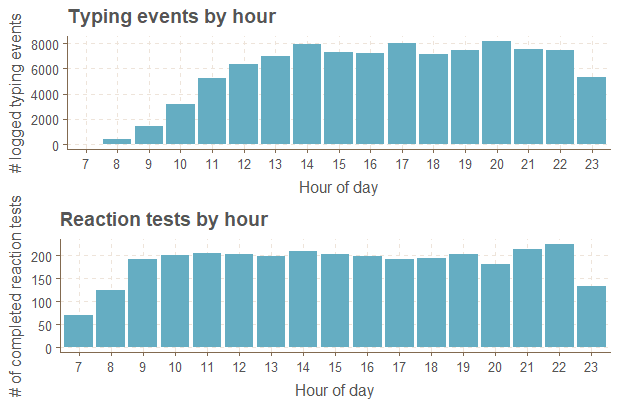}
    \caption{Hourly sum of our typing event dataset and reaction test dataset. Samples between midnight and 7AM (24-06) are omitted from further analysis.}
    \label{fig:hourly_summary}
\end{figure}

The data was preprocessed to remove any abnormal or outlier data, e.g., nights when the participant's nightly heart rate did not drop to the appropriate resting level as a result of, typically, sickness or excessive alcohol intake. We used a threshold of 1.2x participants' median nightly heart rate as the threshold as advised by one of the co-authors, who is an expert in sleep research. 152 nights of sleep data were removed as outliers. These nights are already excluded from previously reported numbers (totalling 3612 nights of data). Figure \ref{fig:hourly_summary} shows the distribution of samples from the reaction test and typing datasets, and Table \ref{tab:data_summary} shows the number of data points for each of the datasets or joined datasets. Typing activity increases during the early morning periods and plateaus after noon, while reaction tests are evenly balanced throughout the daytime (9 a.m. to 10 p.m.). These distributions build a solid foundation for analysing the datasets over the day (from morning to evening).

\subsubsection{Reaction Test Data}
Concerning cognitive performance, 84 participants provided sufficient amount of data for evaluation by completing at least two PVT tests on a minimum of 14 unique days. On average, participants finished 2.72 tests daily, with a maximum of seven tests during a day. The tests were evenly distributed during the day (from 7:00 to 23:00), as shown in Figure \ref{fig:hourly_summary}.

\label{results1}
As described above, we use the PVT to measure alertness. The result of such a test task is an array of reaction times (in milliseconds) from the participant during the test. Each test task's reaction times are then averaged to a single numerical value per test: mean reaction time (RT). RT has the highest effect size for assessing alertness via PVT \cite{basner2011maximizing}. We start the analysis by calculating personalised mean RT, variance, and an average number of errors for each participant (lapses). We similarly calculate personalised averages for all typing and sleep-related variables, where applicable. We merge the sleep dataset with the reaction test data according to the participant ID and the test date, giving us a dataset with reaction test results and the sleep data from the previous night. This dataset contains 3247 entries from 76 unique participants. 

\subsubsection{Typing Data}
Seven participants suffered from technical problems with keyboard data, and their data could not be retained or retrieved for analysis. The remaining 79 participants provided 6.3M keyboard events parsed into \textit{typing sessions}. A typing session is a chain of keyboard events with no longer than two seconds between two actions (typing a character, removing or inserting text). Each typing session consists of mean typing speed (time between events), median typing speed, variance and standard deviation of typing speed, total typing event duration, and length of typed characters. We used a maximum string length of 300 to remove outlier events, such as participants simply pasting long text snippets. The finalised typing dataset contains 234.5K typing events from 79 unique participants.

The keyboard sensor requires specific system access on the participant's smartphone, called the accessibility option. Android vendors implement versions of Android OS, which can cause system options like accessibility or battery management options to behave differently on different vendors' models. During our study and previous similar studies, we have recognised occasional breaks in data collection due to the accessibility option being reset or the participants' smartphone taking an excessively aggressive battery management stance, causing the background logging to cease momentarily.

These breaks can cause a pause in the \textit{continuous} tracking of typing or can cause the reaction test results to be from a period of time (e.g., two weeks) where no tests are finished on each day - i.e., there are days in-between when no tests are finished. As seen in Figure \ref{fig:hourly_summary}, the unavoidable breaks in the datasets have not caused any specific hourly bias. Thus, any 'gaps' in the collected dataset do not influence our analysis.

Next, we look for insights into sleep-related metrics that influence cognitive performance, analyse the relationship between sleep quality and typing metrics, and then model cognitive performance through smartphone keyboard typing behaviour from data collected in real-world situations.

\begin{table}
\small
\begin{tabular}{l|ccl}\\
\textbf{Datatype} &
  \textbf{\# of users} &
  \textbf{Datapoints} &
  \textbf{Description} \\ \hline
Sleep data      & 86 & 3764   &  \begin{tabular}[c]{@{}l@{}}Sleep metrics, collected through the\\device vendor's cloud service\end{tabular} \\
Reaction test & 84 & 3956   &  \begin{tabular}[c]{@{}l@{}}The PVT cognitive measurement test \\performed by the participants\end{tabular} \\
Typing data  & 79 & 234.5K &  \begin{tabular}[c]{@{}l@{}}Typing events collected through the \\AWARE framework keyboard sensor\end{tabular} \\
Reaction+Sleep &
  76 &
  3247 &
  \begin{tabular}[c]{@{}l@{}}Previous nights' sleep quality included \\ for each individual test result.\end{tabular}
\end{tabular}
\caption{Summary of the collected and merged datasets.}
\label{tab:data_summary}
\end{table}

\subsection{Leveraging Individual Sleep Quality Metrics}
\label{results2}

Next, we leverage the detailed sleep metrics available through the wearable ring, how factors like REM sleep or sleep latency impact one's alertness. We investigate whether a single night's sleep metrics impact alertness and whether sleep metrics averaged over the previous five days have a similar or contrasting effect. For each sleep metric, we calculate a 5-day rolling average.

Alertness measured through the PVT follows a declining trend throughout the day due to homeostatic pressure \cite{burke2015sleep}, and with typical peaks (performance drops) at noon and towards the evening and dips (performance increases) during the afternoon caused by circadian rhythm and daily activities such as dining \cite{tag2019continuous}. We model the expected behaviour of the test result in our population by first building a regression model with a standard error of $<$ .001 \revision{seconds} (p $<$ .005) that shows the expected PVT result based on the time of day\revision{, e.g., for P46 at 3 PM the mean reaction time could be 310ms}. We can then calculate a difference for each PVT result (performed at a specific time of the day) and see whether the participant performed on this PVT test better or worse than modelled. \revision{For example, on day four, P46 could perform a reaction test with a mean reaction time of 298ms, indicating a performance of 12 milliseconds better than average. This value is henceforth referred to as the \textit{PVT result difference}. With the regression model, we have a statistically accurate model of reaction test performance throughout the day.} Finally, we can analyse the combination of individual sleep metrics and each metric separately for its effect on the expected PVT result.

\begin{table}
\tiny
\begin{tabular}{l|lcc|lcc}
                                                                    & \multicolumn{3}{c|}{\textbf{One-day sleep metrics}}                                                                                                                                                                                                                                                                     & \multicolumn{3}{c}{\textbf{5-day rolling average}}                                                                                                                                                                                                                                                                    \\
                                                                    & \textbf{Best fit}                                                            & \textbf{Significance}                                                                                                                     & \textbf{\begin{tabular}[c]{@{}c@{}}Coefficient\\ effect size\end{tabular}}                   & \textbf{Best fit}                                                          & \textbf{Significance}                                                                                                                     & \textbf{\begin{tabular}[c]{@{}c@{}}Coefficient\\ effect size\end{tabular}}                   \\ \hline
\textbf{Score}                                                      & \multicolumn{1}{l|}{\begin{tabular}[c]{@{}l@{}}Stepwise\\ LM\end{tabular}}   & \multicolumn{1}{c|}{\begin{tabular}[c]{@{}c@{}}p \textless .0005, \textless{}69\\ p \textless .0005, \textgreater{}69\end{tabular}}       & \begin{tabular}[c]{@{}c@{}}-.187, \textless{}69\\ +.059, \textgreater{}69\end{tabular}       & \multicolumn{1}{l|}{\begin{tabular}[c]{@{}l@{}}Stepwise\\ LM\end{tabular}} & \multicolumn{1}{c|}{\begin{tabular}[c]{@{}c@{}}p \textless .0005, \textless{}74\\ p \textless .05, \textgreater{}74\end{tabular}}         & \begin{tabular}[c]{@{}c@{}}-.189, \textless{}74\\ +.231, \textgreater{}74\end{tabular}       \\
\textbf{\begin{tabular}[c]{@{}l@{}}Total\\ sleep\end{tabular}} & \multicolumn{1}{l|}{\begin{tabular}[c]{@{}l@{}}Stepwise\\ LM\end{tabular}}   & \multicolumn{1}{c|}{\begin{tabular}[c]{@{}c@{}}p \textless .0005, \textless{}6h56m\\ p \textless .0005, \textgreater{}6h56m\end{tabular}} & \begin{tabular}[c]{@{}c@{}}-.148, \textless{}6h56m\\ +.438, \textgreater{}6h56m\end{tabular} & \multicolumn{1}{l|}{\begin{tabular}[c]{@{}l@{}}Stepwise\\ LM\end{tabular}} & \multicolumn{1}{c|}{\begin{tabular}[c]{@{}c@{}}p \textless .0005, \textless 7h\\ p \textgreater .0005, \textgreater 7h\end{tabular}}      & \begin{tabular}[c]{@{}c@{}}-.241, \textless{}7h\\ +.253, \textgreater{}7h\end{tabular}       \\
\textbf{Awake}                                                      & \multicolumn{1}{l|}{\begin{tabular}[c]{@{}l@{}}Multiple \\ GAM\end{tabular}} & \multicolumn{1}{c|}{p \textless .05}                                                                                                      & +.082                                                                                        & \multicolumn{1}{l|}{LM}                                                    & \multicolumn{1}{c|}{p \textless .0005}                                                                                                    & +.075                                                                                        \\
\textbf{Light}                                                      & \multicolumn{1}{l|}{\begin{tabular}[c]{@{}l@{}}Multiple \\ LM\end{tabular}}  & \multicolumn{1}{c|}{p \textless .005}                                                                                                     & -.123                                                                                        & \multicolumn{1}{l|}{LM}                                                    & \multicolumn{1}{c|}{p \textless .0005}                                                                                                    & -.073                                                                                        \\
\textbf{REM}                                                        & \multicolumn{1}{l|}{\begin{tabular}[c]{@{}l@{}}Multiple \\ LM\end{tabular}}  & \multicolumn{1}{c|}{p \textless .05}                                                                                                      & -.078                                                                                        & \multicolumn{1}{l|}{\begin{tabular}[c]{@{}l@{}}Stepwise\\ LM\end{tabular}} & \multicolumn{1}{c|}{\begin{tabular}[c]{@{}c@{}}p \textless .05, \textless{}1h20m\\ p \textless .05, \textgreater{}1h20m\end{tabular}}     & \begin{tabular}[c]{@{}c@{}}+.140, \textless{}1h20m\\ -.109, \textgreater{}1h20m\end{tabular} \\
\textbf{Deep}                                                       & \multicolumn{1}{l|}{LM}                                                      & \multicolumn{1}{c|}{p = 0.63}                                                                                                             & -                                                                                            & \multicolumn{1}{l|}{LM}                                                    & \multicolumn{1}{c|}{p = 0.338}                                                                                                            & -                                                                                            \\
\textbf{\begin{tabular}[c]{@{}l@{}}Bedtime \\ duration\end{tabular}}    & \multicolumn{1}{l|}{\begin{tabular}[c]{@{}l@{}}Stepwise\\ LM\end{tabular}}   & \multicolumn{1}{c|}{\begin{tabular}[c]{@{}c@{}}p \textless .0005, \textless{}8h33m\\ p \textless .0005, \textgreater{}8h33m\end{tabular}} & \begin{tabular}[c]{@{}c@{}}-.304, \textless{}8h33m\\ +.289, \textgreater{}8h33m\end{tabular} & \multicolumn{1}{l|}{\begin{tabular}[c]{@{}l@{}}Stepwise\\ LM\end{tabular}} & \multicolumn{1}{c|}{\begin{tabular}[c]{@{}c@{}}p \textless .0005, \textless{}8h37m\\ p \textless .0005, \textgreater{}8h37m\end{tabular}} & \begin{tabular}[c]{@{}c@{}}-.309, \textless{}8h37m\\ +.216, \textgreater{}8h37m\end{tabular} \\
\textbf{\begin{tabular}[c]{@{}l@{}}Sleep\\ latency\end{tabular}}    & \multicolumn{1}{l|}{\begin{tabular}[c]{@{}l@{}}Multiple\\ LM\end{tabular}}   & \multicolumn{1}{c|}{p \textless .0005}                                                                                                    & +.141                                                                                        & \multicolumn{1}{l|}{\begin{tabular}[c]{@{}l@{}}Multiple\\ LM\end{tabular}} & \multicolumn{1}{c|}{p \textless .0005}                                                                                                    & +.200                                                                                        \\
\textbf{Restless}                                                   & \multicolumn{1}{l|}{LM}                                                      & \multicolumn{1}{c|}{p = 0.07}                                                                                                             & -.034                                                                                        & \multicolumn{1}{l|}{\begin{tabular}[c]{@{}l@{}}Multiple\\ LM\end{tabular}} & \multicolumn{1}{c|}{p \textless .05}                                                                                                      & -.057                                                                                        \\
\textbf{\begin{tabular}[c]{@{}l@{}}Lowest \\ HR\end{tabular}}       & \multicolumn{1}{l|}{\begin{tabular}[c]{@{}l@{}}Multiple\\ LM\end{tabular}}   & \multicolumn{1}{c|}{p \textless .005}                                                                                                     & +.090                                                                                        & \multicolumn{1}{l|}{\begin{tabular}[c]{@{}l@{}}Multiple\\ LM\end{tabular}} & \multicolumn{1}{c|}{p \textless 0.005}                                                                                                    & +.093                                                                                        \\
\textbf{RMSSD}                                                      & \multicolumn{1}{l|}{\begin{tabular}[c]{@{}l@{}}Multiple\\ LM\end{tabular}}   & \multicolumn{1}{c|}{p \textless .05}                                                                                                      & -.068                                                                                        & \multicolumn{1}{l|}{\begin{tabular}[c]{@{}l@{}}Multiple\\ LM\end{tabular}} & \multicolumn{1}{c|}{p \textless 0.05}                                                                                                     & -.092                                                                                       
\end{tabular}
\caption{Best fitting regression models for each of the sleep metrics as inputs and the PVT result difference as output. Tests are from one-day analysis of the previous night's data (left columns), and the rolling average from previous five days (right columns). LM indicates Linear Regression Model and GAM indicates Generalized Additive Model. Coefficient column does not include the base coefficient a from the regression equation: L = a + $b_1$X + $b_2$Y ..., only the coefficient sizes of the sleep metrics ($b_1$, $b_2$, etc.). Due to repeated tests, the significance levels should be considered with Bonferroni corrections, with p $<$ .05 being borderline significant.}
\label{table:models}
\end{table}

\subsubsection{Previous Night's Effect on Alertness}
\revision{In the following section we provide regression models to estimate the effect of different sleep metrics on the \textit{PVT result difference}. Depending on the fit of the model used, different sleep metrics have best estimates with varying models chosen according to statistical significance. Full summary of best fitting models is shown in Table \ref{table:models}.}

To evaluate the effect of each sleep metric on PVT result difference, we start by building a multiple linear regression model from the 11 metrics: \textit{sleep score, bedtime duration, awake, light, REM, deep, total sleep, sleep latency, restless periods, lowest HR, rmssd} \revision{as input variables and the \textit{PVT result difference} as the result variable}. The model has a combined coefficient estimate of .454 (normalised) with p $<$ .0005. Two variables (\textit{restless, awake}) are non-significant in the model\revision{, meaning they have no relation to the output variable.} Two variables (\textit{deep sleep, total sleep}) \revision{have poor fit} as coefficients in the model as they are linearly related to other variables. This is understandable as \textit{total sleep time} is both intuitively and statistically (p $<$ .0005, cor=.90 using Pearson's Correlation) correlated with \textit{bedtime duration}. For \textit{deep sleep}, the largest correlation effect is with \textit{light sleep} (p $<$ .005, cor=-.37). 

To analyse and model the non-significant variables from the multiple linear model (\textit{restless, awake, deep sleep, total sleep}), we build a multiple generalised additive regression model (GAM) using the same set of 11 metrics. GAM is a linear regression model that applies a smoothing function to the data, effectively creating a smoothed model of the data. \textit{Awake} was found to be significant in GAM but only at p = .03. Rest of the four variables were non-significant. As part of the GAM, we visually inspect each of the variables' effect on the PVT result and notice that a subset of the variables (\textit{score, total, duration}) follow a V-shaped curve. For these variables, we investigate the optimal breakpoint between decline and increase (or increase and decline) to better model the variable's effect on the PVT result. We find the breakpoints to be 69 for \textit{sleep score}, 8h33m for \textit{bedtime duration}, and 6h56m for \textit{total sleep}. We then validate these breakpoints by ensuring that the linear regression model is statistically significant before and after this breakpoint (p $<$ .005 for all cases). 

As we have conducted multiple regression tests (both linear and non-linear), according to recommendations from, e.g., Li et al. \cite{li2017introduction}, the significance level is adjusted to negate the potential increase in type I errors due to repeated tests. Thus, $\alpha$ = 5\% would not be considered significant. As seen in Table \ref{table:models}, the majority of the test results are significant at $\alpha$ = 1\% or lower (a limit used with Bonferroni corrections\cite{weisstein2004bonferroni} with five repeated tests per sample). Thus, we can conclude that this analysis has no increased probability of false positives.

\subsubsection{Effect of 5-night Average on Alertness}
We then repeat the same procedure after calculating a rolling 5-day average of all sleep metrics from previous days' nights. The initial multiple linear regression reveals only four variables (\textit{sleep latency, restless, lowest HR, rmssd}) as significant. GAM shows the same four variables as significant. Continuing with visual inspection and analysing each remaining variable with individual linear regression models, \textit{score, awake time, light sleep}, and \textit{total sleep} follow a linear model. \textit{Sleep score}, \textit{total sleep, duration} and \textit{REM sleep} follow a stepwise linear function with breakpoints at 74 (\textit{sleep score}), 8h37m (\textit{bedtime duration}), 1h20m (\textit{REM}), and 7h (\textit{total sleep}). 

\begin{figure} [h]
    \centering
    \includegraphics[width=1\columnwidth]{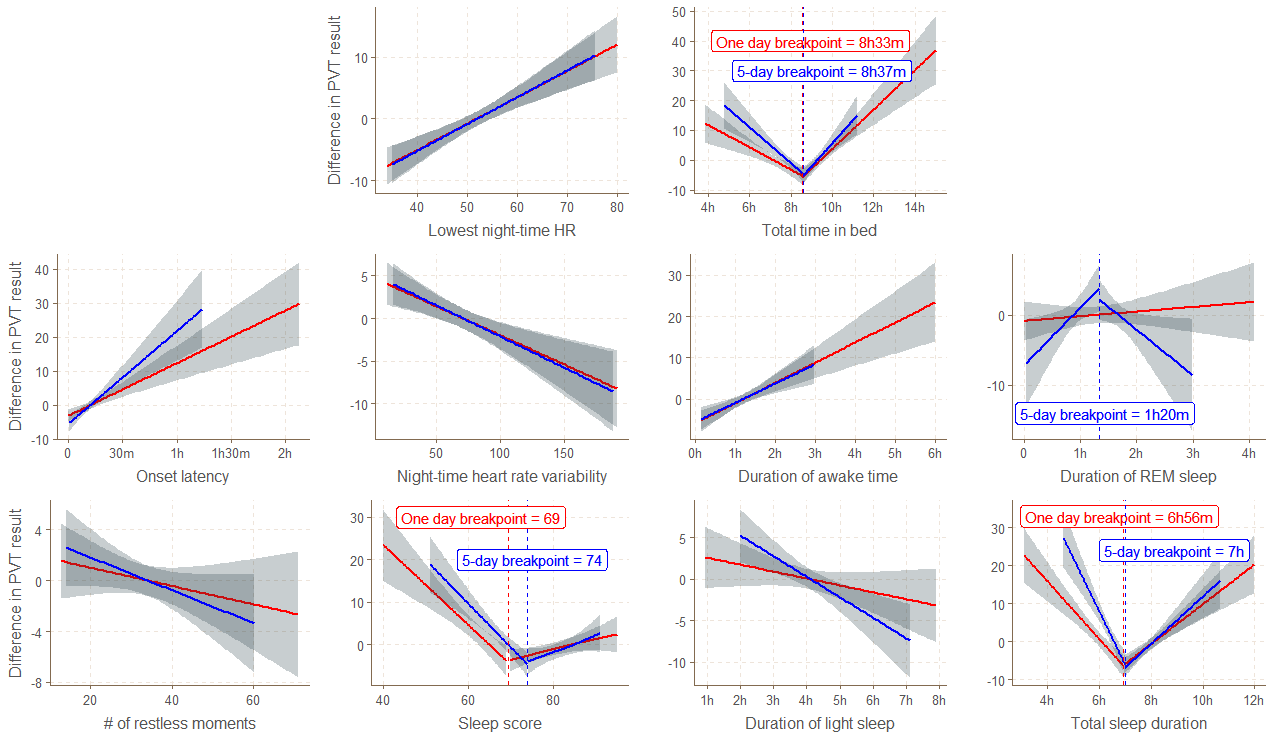}
    \caption{Regression model results between reaction test performance (difference to mean, lower value is better) and individual sleep metrics. Blue line denotes a rolling average from previous five nights for each sleep metric. Red line denotes previous night's result. Grey area denotes the 95\% confidence interval of the model. Table \ref{table:models} includes details on model selection for each metric and one-day vs. 5-day results. 'Total time in bed' equals the \textit{duration} variable and 'Total sleep duration' equals the \textit{total sleep} variable.}
    \label{fig:sleep_metrics}
\end{figure}

We have summarised the coefficients and significance of each best-performing model for each sleep metric in Table \ref{table:models} and visualised the effect of each sleep metric (excluding deep sleep as it had no statistically significant results) in Figure \ref{fig:sleep_metrics}. In the following subsections, we will describe the effect and effect size of each of the sleep metrics on the performance change in the PVT reaction test according to our analysis.

\subsubsection{Sleep score:} For both the previous night and the 5-day average, sleep scores below a breaking point of 69 (for the previous night) and 74 (for the 5-day average) show a sharp decline in PVT result. Higher sleep scores show a slight decline, but the base coefficient is relatively low in both cases, keeping the effect to only a few milliseconds in the PVT result (e.g., an increase from -4 to +4) over an increase of up to 20 points in sleep score (e.g., from 74 to 94), as can also be seen in Figure \ref{fig:sleep_metrics}. Overall, it is clear that suboptimal sleep very rapidly shows a decline in the following day(s) performance. The Oura ring sleep algorithm categorises any sleep below a sleep score 70 as bad sleep.

\subsubsection{Total sleep duration \& Total time in bed:} Similar to sleep score, both \textit{total sleep} and \textit{bedtime duration} show a sweet spot where the PVT performance is best, with the PVT test performance declining when spending too little or too much time in bed, or too little or too much time in actual sleep. On average, the participants spent 1h10min awake in bed before or after the actual sleep period or awake at night.
\subsubsection{Sleep phases:} The four phases of sleep (\textit{REM, light, deep, awake}) each have a separate function during sleep, and three out of the four show distinct effect on the PVT performance. Less \textit{awake} time shows improvements in PVT performance for both 1-day and 5-day sleep behaviour - with less than one hour of awake time before, after, or during the night can improve your performance. An increase in \textit{light sleep} also shows an improvement in PVT performance, with more \textit{light sleep} in the 5-day average showing significantly (p $<$ .05) larger benefits. Lastly, for \textit{REM sleep}, there is a very slight decline in 1-day performance as the duration in REM sleep increases, and for the 5-day average, we find a suboptimal point of 1h20m of \textit{REM sleep} during the night, with the PVT performance increasing both before and after this point.

\subsubsection{Heart-rate:} \textit{HRV} and \textit{lowest heart rate} are both important for improving reaction test results. Lower night-time HR and higher HRV indicate a better overall state in the body and strongly impact alertness the following day. There are no differences for 1-day or 5-day behaviour, likely explained by the metrics being based on an individual's physiological attributes. Thus, the metrics do not show much within-subject variance across different nights.
\subsubsection{Sleep latency \& restfulness:} Difficulties in falling asleep unanimously show declining alertness for both the previous night and the 5-night average. The 5-night average has a significantly stronger influence on the decline, with close to double the rate. High \textit{sleep latency} can also negatively impact solely due to the amount of actual sleep being diminished by its influence - presumably, a person has to wake up at a specific time regardless of how long it takes to fall asleep. A higher number of \textit{restless} events imply better alertness the following day. However, the significance of this finding is non-significant when only considering the previous night, and the actual effect size is the smallest of small sleep metrics, with the average only varying between +2ms to -2ms.

\subsubsection{Summary}
We have shown that different detailed sleep metrics have varying impacts on a cognitive performance test. Factors like the overall amount of sleep, the time it takes to fall asleep and more physiological factors like heart rate and heart rate variability strongly impact our alertness. We also highlight how short-term (one night) benefits differ from longer-term (5-night average) sleep habits.

\subsection{Reaction Test Performance and Typing}
\label{results3}
Cognitive computing aims to study cognitive performance through datasets collected from humans. However, datasets used as baseline are limited to inputs from specific applications or tests conducted by study participants. Cognition-aware systems are based on datasets that can mimic the behaviour of these tests, and such systems attempt to model human cognitive performance through external variables to reduce the reliance on explicit test cases - which often limit the duration of in-the-wild experiments. Ubiquitous computing can be leveraged to create systems based on, e.g., mobile sensing, that can model cognitive performance through external datasets, e.g., mobile game performance \cite{dingler2020extracting}, wearable technologies \cite{tag2020inferring}, or external sensors \cite{abdelrahman2017cognitive}. Considering our study design, we can observe that passively collecting datasets, e.g., typing data, can be easily collected over a longer period. Our study setting enabled the PVT test for two weeks, typically the upper limit for effectively using the Experience Sampling Method \cite{van2017experience}. 

The concluding section of our work aims to mimic human cognitive performance and measurements by analysing typing behaviour. Namely, we look at typing speed and proneness to mistype or perform cognitive errors that require backtracking. These factors were selected as they likely best mimic the cognitive performance measurement metrics of the PVT test. Our dataset contains 101K typing events matched with the previous night's sleep data and 3182 reaction test results, again matched with the previous night's sleep data. Similar to the reaction test data, we use \textit{time awake} (in hours) as the metric for the passage of time. As each reaction test and each typing event is performed at a specific time of day, we can, to some extent, match events that occur concurrently (on the same day). In our dataset, 4202 typing events occur within one hour of a concurrent reaction test, and these can be considered \textbf{matched samples}. In contrast, the \textbf{generic samples} simply match reaction tests and typing events based on time spent awake, regardless of the date. These matched samples have an average time difference of 25m 57s (SD = 17m 51s) between the two and follow a long-tailed distribution with a peak at 5m 22s (with non-normal distribution verified using Shapiro-Wilk test with p $<$ .005 and W = 0.93). Similarly, looking at quantiles of the time differences, 50\% of the samples are within 23m 30s, and 25\% are within six minutes. The PVT is sensitive to circadian effects, so we wanted to ensure that the matched samples are occurring relatively close to each other, which this analysis verifies.

Typing speed is measured as the time (in milliseconds) between two keystrokes and is divided into sessions at least two seconds apart. The dataset's average time between two keystrokes is 329.9ms (SD = 140.8, IQR = 240-383). Like the reaction test result, the typing speed follows a declining linear trend, with identifiable performance surges before and after six hours after waking up - typically around noon or early afternoon. Using Pearson's correlation, we can observe a significant (but weak) linear correlation between typing speed to waketime (p $<$ .05, r = .02) and typing speed to corrections (p $<$ .05, r = -.07).

\subsubsection{Modelling Reaction Test Performance from Typing Metrics}
Next, we show that typing speed can be used as a proxy for cognitive performance, as measured by the PVT test. We use the same approach for both the \textbf{generic samples} (events that occur after similar \textit{waketimes} regardless of the day) and \textbf{matched samples} (events that occur precisely one hour from each other during the same day). To better understand the similarity between the PVT reaction test data and the typing characteristics, we rely on the similarity between the daily curvature. Using the \textit{stat\_smooth} R function with the GAM (generalized additive model) smoothing function, we can interpolate both the generic and matched samples. For the interpolation function, the 95\% confidence interval is well within one standard deviation for both the reaction test and typing data and for both generic and matched samples. This implies that the interpolation results are statistically meaningful. The details are shown in Table \ref{tab:interpolationci}. 

\begin{figure}[h]
    \centering
    \includegraphics[width=0.7\columnwidth]{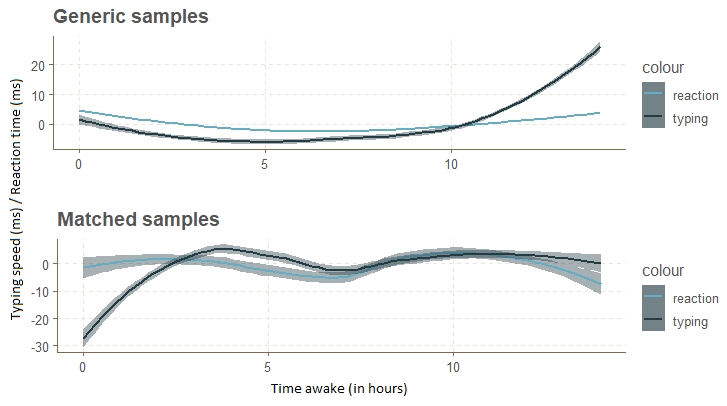}
    \caption{Comparison of typing speed (milliseconds between keystrokes) to reaction test result (in milliseconds) over the course of the day according to time spent awake (in hours), using the GAM smoothing function implementation from the 'ggplot2' R library. Both generic samples and matched samples show strong similarities.}
    \label{fig:interpolated_comparison}
\end{figure}

\begin{table}
\centering
\tiny
\begin{tabular}{l|cc}
                                     & \textbf{Origin data SD}   & \textbf{Interpolation 95\% CI}         \\ \hline
\begin{tabular}[c]{@{}l@{}}\textbf{Reaction test (generic samples interpolation)}\end{tabular} & 43.6  & 333.5 - 339.3  \\
\begin{tabular}[c]{@{}l@{}}\textbf{Reaction test (matched samples interpolation)}\end{tabular}  & 43.6  & 321.0 - 327.9  \\
\begin{tabular}[c]{@{}l@{}}\textbf{Typing speed (generic samples interpolation)}\end{tabular}  & 140.9 & 328.9 - 334.0  \\
\begin{tabular}[c]{@{}l@{}}\textbf{Typing speed (matched samples interpolation)}\end{tabular}   & 140.9 & 319.0 - 340.6 
\end{tabular}
\caption{Interpolation results.}
\label{tab:interpolationci}
\end{table}

Figure \ref{fig:interpolated_comparison} shows the curvature between reaction test results and typing speed. The similarity of the curve is apparent for both generic samples and especially matched samples, as the circadian process \cite{van2000circadian} can be observed as slight performance increases after being awake for 4-7 hours. Using the Kolmogorov-Smirnov test, we can verify the similarity between typing speed and cognitive performance for both generic samples (p $<$ .005, D = .55) and paired samples (p $<$ .005, D = .8125). Lastly, we use the Bland-Altman method to ensure that the pairwise comparison in the interpolated data is significantly similar. Also known as the 'Tukey mean-difference plot', the method is commonly used to compare two measurement methods for devices that measure the results of the same sample(s), in our case typing speed to the baseline (PVT).



\subsection{Summary}
Relying on constant cognitive tests is not a reasonable way to collect long-term data about cognitive performance, as participant motivation significantly declines after the initial participant weeks \cite{van2017experience}. Previous work has analysed various methods for collecting such data unobtrusively, and we extend these works by showcasing how smartphone typing speed functions as a proxy for cognitive performance measurements. By its ubiquitous nature, smartphone typing could offer a tool for continuous and long-term cognitive measurements.

\section{Discussion}
The current trend of including sleep-tracking algorithms in activity trackers and other wearable devices offers potential for cognitive and attention-computing research. Given how sleep is one of the main driving factors behind our physiological and mental well-being, these devices' detailed sleep quality metrics offer tremendous potential for self-reflection and learning about our sleeping habits in general. While it is undeniable that these devices - although validated for accuracy by CS researchers \cite{de2018validation,de2015validation} - are perhaps not yet perfectly suited for \textit{clinical} research, the devices are clearly helpful for those using them and thus for the human-centric HCI field as well and their use has recently picked up for research focuses \cite{perez2020future}. To the best of our knowledge, the study contributed in this article is one of such early attempts to capture human behaviour on a large scale by using such digital sleep trackers. 

Using either researcher-provided or self-owned wearable devices for \textit{in-the-wild} studies has its challenges. Wearable abandonment and lack of continuous use is a well-known challenge \cite{shin2019wearable,lee2017look}. While battery life in wearable devices has significantly improved over the past five years, data gaps are still a frequent problem, especially for study participants or users who are not well self-motivated or lack an inherent interest in self-tracking. The novelty effect can also wear off after the initial few weeks, causing data gaps later in a study setting. While our study participants were keenly interested in trying out a new novel piece of technology, we did not observe a significant decline in data quantity over the study period, with typical gaps in sleep tracking lasting only for one or two nights. Based on the success of the Oura Ring, we hypothesise that the ring form factor is both convenient and compelling. Using the ring allowed us to collect a high-quality dataset without many gaps: Continuous data from most participants across a two-month timespan. Indeed, only 17.9\% of the reaction test entries were not associated with the previous night's sleep information. 

Cognitive performance plays a key role in HCI topics adjacent to cognitive-aware computing, such as attention management, design of notification delivery systems, and health and wellbeing-related topics. Thus, understanding more about underlying influences in cognitive performance can be helpful for future research in these fields. Our results show how sleep interacts with cognitive performance, specifically alertness, as that is what the PVT test mainly measures, and the findings can leverage further in encouraging or teaching people to sleep better. For example, many self-tracking communities are aware of both mechanisms and \textit{reasons} for improving one's sleep, but many positive effects are still either neglected or not part of the public awareness. Stronger public dissemination of these associations, e.g., our results, could help individuals motivate themselves to improve their sleep. This could benefit the ageing population in particular, as, e.g., the decline in non-REM sleep duration can indicate worsening cognitive decline \cite{taillard2019non}. Furthermore, decline in sleep quality shows a parallel trajectory to cognitive and physical health decline in older adults \cite{djonlagic2021macro}, indicating that earlier interventions in sleep hygiene and sleep quality could delay such decline.

Althoff et al. \cite{althoff2017online} analysed the impact of subsequent insufficient nights of sleep on cognitive performance and released similar to our findings that poor sleep can have an impact of 8-10ms on reaction time - and also that after five-to-six night nights of insufficient sleep the body adapts and the performance returns to normal levels. Our results from looking at five-night averages reveal similar differences in numerous sleep metrics. One curious difference between the previous to five-night average is that the amount of REM sleep and total sleep duration is very selective, and either too much or too little will negatively impact cognitive performance. The lack of sleep causing worse reactions is rather intuitive, but too much sleep's negative effect warrants further discussion. One explanation could be that "oversleeping" is likely a result of a prior lack of sleep, which in turn causes fatigue and poor performance. Another reason could be due to situations like being sick, being highly stressed, or simply being already fatigued from overworking. Regardless, Herscovitch et al. \cite{herscovitch1980changes} revealed that both sleep deprivation and recovery oversleeping cause a decline in cognitive performance.

Finally, the HCI community has for years been interested in developing novel proxies for human behaviours or functions through exploring different tracking devices and methods. Such proxies are at the heart of \textit{ubiquitous} tracking through everyday devices. For cognitive performance, different types of games embedded with cognitive measurements have been explored \cite{dingler2020extracting}, and associations have been drawn from (limited) desktop typing activities to sleep quality \cite{althoff2017harnessing}. But smartphones have accounted for the majority of our technology and internet usage ever since 2016\footnote{https://techcrunch.com/2016/11/01/mobile-internet-use-passes-desktop-for-the-first-time-study-finds/?guccounter=1}. We spend significant amounts of time on smartphones and the vast majority of our non-verbal communication occurs via the smartphone. Vizer \cite{vizer2009automated} showed that stress detection can be automated from variance in keystroke and linguistic features, showcasing that changes in cognitive stress levels are measurable from typing data. In our case, we are not identifying any specific behaviour from the typing data but rather showcasing how typing data \textit{matches} the cognitive performance of the PVT test. Thus, analysing the typing behaviour on our smartphones can offer rich insights into our mental and cognitive state \textit{throughout the day}. There are drawbacks, naturally, as users can perceive keylogging and similar methods as intrusive, and applications designed to track typing behaviour require extra permissions to function appropriately. Large open datasets of typing behaviours would undoubtedly raise eyebrows, at least if they were not appropriately obfuscated and anonymised. However, as part of our findings shows, merely looking at typing \textit{metrics} can offer important insights about individuals.

\subsection{Limitations}
We acknowledge the following limitations in our work. Any study design reliant on mobile sensing exhibits some data loss, which is unfortunate but often unavoidable. This applied to both our wearable device data as well as mobile data. These have been accounted for, as explained in the data analysis section. Additionally, the ESM approach typically includes a loss of participant retention over the study period, which we accounted for by increasing the sampling quantity. This leads to a potential decline in data quality as the PVT test is not as sensitive in capturing the effect of poor sleep when conducted as a short-form test of two minutes or less, compared to the standardized 5-minute or 10-minute PVT tests.

\section{Conclusion}
This paper presents the results of a two-month-long data collection from digital sleep-tracking devices in combination with smartphone-based reaction test data from 14 study days and typing metrics collected throughout the study. We showcase how detailed sleep metrics can offer insights about cognitive performance beyond what was previously known. Differences in sleep phases and sleep metrics impact cognitive performance; physiological measurements like night-time heart rate and HRV, the overall amount of sleep, and sleep latency strongly influence a cognitive performance test. We also showcase how smartphone typing could be efficiently leveraged as a proxy for cognitive performance tests in future studies and, as such, could enable longer studies that would be less disruptive in the participants' daily lives.

\subsection{Acknowledgements}
This research is connected to the GenZ strategic profiling project at the University of Oulu, supported by the Academy of Finland (project number 318930), and CRITICAL (Academy of Finland Strategic Research, 335729). Part of the work was also carried out with the support of Biocenter Oulu, spearhead project ICON. 

\section{Conflicts of Interests}
On behalf of all authors, the corresponding author states that there is no conflict of interest.

\section{Data Availability}
The data produced as part of this work will be made openly available on request.


 \bibliographystyle{elsarticle-num} 
 \bibliography{cas-refs}





\end{document}